\documentclass[12pt]{article}
\usepackage{amsbsy}
\begin{document}
\begin{center}
{\bf\Large Hamilton-Jakobi method for classical           
       mechanics in Grassmann algebra}
\end{center}
\centerline{Kyrylo V. Tabunshchyk}                  
\begin{center}
 Institute for Condensed Matter Physics\\
 of the National Academy of Sciences of Ukraine,\\
 1~Svientsitskii St., 79011 Lviv, Ukraine
\end{center}

\begin{abstract}
 {We present the Hamilton-Jakobi method for the
classical mechanics with constrains in Grassmann algebra.
 In the frame of this method the solution for the classical system
characterized by the SUSY Lagrangian is obtained.}
\end{abstract}

Key words: Hamilton-Jakobi method, SUSY system, classical
mechanics, constrains.

PACS number(s): 03.20, 11.30.P

\def\rz{\mbox{${\sf I \! R}$}}\

  The problem of Lagrangian and Hamiltonian mechanics with
Grassmann variables has been discussed previously in works
\cite{Junker1,Junker2,Junker3} and an examples of solutions for
classical systems were presented.

 In this paper we propose the Hamilton-Jakobi method for the solution of the
classical counterpart of Witten`s model \cite{Witten}.

 We assume that the states of mechanical system are described by
the set of ordinary bosonic degrees of freedom $q$ (even Grassmann
numbers) and the set of fermionic degrees of freedom $\psi$ (odd
Grassmann numbers).

 The Hamilton-Jakobi equation in the case of the classical mechanics with
constrains in Grassmann algebra is following:
\begin{equation}
 \label{equation1}
     \frac{\partial S}{\partial t}=
     - H\bigg(q,\psi ,\frac{\partial S}{\partial q},
                      \frac{\partial S}{\partial\psi},
                      \lambda\big(q,\psi,
                      \frac{\partial S}{\partial q},
                      \frac{\partial S}{\partial\psi}\big),
                      \lambda^\alpha , t\bigg).
\end{equation}
 Here, $\lambda$ is the set of certain Lagrange multipliers for the
constrains which can be found from the equations of motion and
from the time-independence conditions.
 On the other hand, $\lambda^\alpha$ are those
multipliers which cannot be found and which form a functional
arbitration for solutions (in the theory with the first-class
constrains \cite{Hojman,Dirac,Tyutin}).
 However, we can transform the theory with the
first-class constrains to the physically equivalent theory with
the second-class constrains. As an example, we can take the strong
minimal gauge which does not shift the equations of motion (the
so-called canonical gauge $G^{(c)}$ \cite{Gitman}).

\centerline{\underline{Jakobi theorem.}}

 Let us consider a full solution of the Hamilton-Jakobi equation
$S{=}S_r(q,\psi,\alpha,\beta,t)$ ($\alpha$ is a set of the even
Grassmann constants, $\beta$ is a set of the odd ones).
 We perform the canonical transformation from the old variables
$q$, $\psi$, $P_q$, $P_{\psi}$ to the new ones (taking $S_r$ as a
generating function) and put $\alpha=P_Q$, $\beta=P_\nu$ as a new
canonical momenta and $Q$, $\nu$ as a new coordinates.
 Then the relations between the new and old variables can be written in the form:
\[
 H'=H+\frac{\partial S_r}{\partial t},\quad
 P_q=\frac{\partial S_r}{\partial q},\quad
 P_\psi=\frac{\partial S_r}{\partial\psi},\quad
 Q=\frac{\partial S_r}{\partial P_Q},\quad
 \nu=-\frac{\partial S_r}{\partial P_\nu}.
\]
 Since $S_r$ is the solution of the Hamilton-Jakobi equation, we
obtain that
\[
 H'=0\quad\Longrightarrow\quad P_Q=\mbox{const},\quad Q=\mbox{const},\quad
 P_\nu=\mbox{const},\quad \nu=\mbox{const},
\]
new coordinates are constant.
 From the obtained result we can write:
\begin{eqnarray}
 \label{Jakobi}
&&\partial S_r /\partial\alpha
                              =\mbox{const (even Grassmann number)},\\
&&\partial S_r /\partial\beta
                              =\mbox{const (odd Grassmann number)}. \nonumber
\end{eqnarray}
 The solution of the equations (\ref{Jakobi}) gives the variables
$q$ and $\psi$ as functions of time.
 Time dependencies of the canonical momenta can be found from the relations
 $P_\psi=\partial S/\partial \psi$,
 $P_q=\partial S/\partial q.$

 Let us now consider the Lagrangian \cite{Junker1,Junker2}
\begin{equation}
 \label{susyl}
 L=\frac{\dot q^2}2-\frac 12V^2(q)
   -\frac{\rm i}2(\dot{\bar\psi}\psi-\bar\psi\dot\psi)
   -U(q)\bar\psi\psi.
\end{equation}
which possess supersymmetry when $U(q)=V'(q)$ (in this case the
real function $V$ is the so-called superpotential) \cite{Witten}.
 The overbar denotes the Grassmann variant of the complex conjugation.

 The momenta conjugate to the fermionic variables do not depend on
$\dot\psi$ or $\dot{\bar\psi}$.
 Hence, we have the following constrains between coordinates and
momenta:
\begin{equation}
 \label{constrains}
 F_1=P_\psi+\frac{\rm i}2{\bar\psi},\qquad
 F_2=P_{\bar\psi}+\frac{\rm i}2\psi.
\end{equation}
 The Lagrange multipliers can be found from the following
time-independence conditions:
\begin{equation}
 \label{conditions}
 \dot{F_1}=\big\{ H(\lambda)\, ,\, F_1\big\}=0,\qquad
 \dot{F_2}=\big\{ H(\lambda)\, ,\, F_2\big\}=0,
\end{equation}
where
\begin{equation}
 H(\lambda)=\frac{P^2_q}2+ \frac 12V^2(q)+ U(q)\bar\psi\psi
            +\lambda_1(P_{\psi}+\frac{\rm i}2{\bar\psi})
            +\lambda_2(P_{\bar\psi}+\frac{\rm i}2\psi).
\end{equation}
 Since the Hamiltonian of the system (\ref{susyl}) is following:
\begin{equation}
 \label{susyh}
  H=H(\lambda(q,\psi,\bar\psi))
    =\frac{P^2_q}2+\frac 12V^2(q)
    -{\rm i}U(q)P_{\bar\psi}\bar\psi+{\rm i}U(q)P_\psi\psi .
\end{equation}
 Starting from the obtained Hamiltonian (\ref{susyh}), the
Hamilton-Jakobi equation (\ref{equation1}) can be written in the
form:
\begin{equation}
\label{equation2}
  \frac{\partial S}{\partial t}
 +\frac 12\left(\frac{\partial S}{\partial q}\right)^2
 +\frac 12V^2(q)
 -{\rm i}U(q)\frac{\partial S}{\partial\bar\psi}\bar\psi+{\rm i}U(q)
 \frac{\partial S} {\partial\psi}\psi=0.
\end{equation}
 Let us make an ansatz for the action
\begin{equation}
 \label{ansatz}
 S(t,q,\psi ,\bar\psi )=S_0(t,q)
                       +\psi\bar\psi S_1(t,q)
                       +\psi S_2(t,q)
                       +\bar\psi S_3(t,q),
\end{equation}
where $S_0$ and $S_1$ are even Grassmann functions and $S_2$ and
$S_3$ are the odd ones.
 After the substitution of ansatz (\ref{ansatz}) into the equation (\ref{equation2})
and decomposition of this equation on Grassmann parities we obtain
the next system of equations:
\begin{eqnarray}
 \label{system}
  \frac{\displaystyle\partial S_0}{\displaystyle\partial t}
 +\frac{\displaystyle 1}{\displaystyle 2}
  \left(\frac{\displaystyle\partial S_0}{\displaystyle\partial q}\right)^2
 +\frac{\displaystyle 1}{\displaystyle 2}V^2(q)=0,\nonumber\\
  \frac{\displaystyle\partial S_2}{\displaystyle\partial t}
 +\frac{\displaystyle\partial S_0}{\displaystyle\partial q}
  \frac{\displaystyle\partial S_2}{\displaystyle\partial q}
 -{\rm i}U(q)S_2=0 ,\nonumber\\
  \frac{\displaystyle\partial S_3}{\displaystyle\partial t}
 +\frac{\displaystyle\partial S_0}{\displaystyle\partial q}
  \frac{\displaystyle\partial S_3}{\displaystyle\partial q}
 +{\rm i}U(q)S_3=0 ,\\
  \frac{\displaystyle\partial S_1}{\displaystyle\partial t}
 +\frac{\displaystyle\partial S_0}{\displaystyle\partial q}
  \frac{\displaystyle\partial S_1}{\displaystyle\partial q}=0,\nonumber\\
  \frac{\displaystyle\partial S_2}{\displaystyle\partial q}
  \frac{\displaystyle \partial S_3}{\displaystyle\partial q}\psi\bar\psi=0.\nonumber
\end{eqnarray}
 The first equation can be integrated by the variable decomposition method.
 Thus, we obtain
\begin{equation}
 \label{S0}
 S_0=\int\!\sqrt{2E-V^2(q)}\,{\rm d}q\, -Et ,
\end{equation}
where $E$ is the constant of integration.
 Starting from the expression (\ref{S0}), for the fourth equation we obtain the
following:
\begin{equation}
 \label{S1}
 S_1=\int\!\frac{A\,{\rm d}q}{\sqrt{2E-V^2(q)}}-At.
\end{equation}
Here, $A$ and $E$ are real variables.

 The solutions of the second and third equations can be written as
\begin{eqnarray}
 \label{S2S3}
  S_2=\phi_1\Bigg(\int\!\frac{{\rm d}q}{\sqrt{2E-V^2(q)}}-t\Bigg)
        \exp\Bigg({\rm i}\int\!\frac{U(q)\,{\rm d}q}{\sqrt{2E-V^2(q)}}\Bigg),\\
        \nonumber
  S_3=\phi_2\Bigg(\int\!\frac{{\rm d}q}{\sqrt{2E-V^2(q)}}-t\Bigg)
        \exp\Bigg(-{\rm i}\int\!\frac{U(q)\,{\rm d}q}{\sqrt{2E-V^2(q)}}\Bigg),
\end{eqnarray}
where $\phi_1$ and $\phi_2$ are arbitrary odd Grassmann functions.
 In our case, it is sufficient to take $\phi_1=const$, $\phi_2=const$.
 The last equation from (\ref{system}) leads to some condition on
functions $\phi_1$ and $\phi_2$, which is satisfied when they are
constant.

 Thus, we can present the action in the following form:
\begin{eqnarray}
 \label{action}
  S&=&\int\!\sqrt{2E-V^2(q)}\,{\rm d}q\,-Et
     +\int\!\frac{\displaystyle A\,{\rm d}q}
     {\displaystyle\sqrt{2E-V^2(q)}}\,\psi\bar\psi
     -At\psi\bar\psi \\
    &+&\psi\phi_1\Bigg(\int\!\frac{\displaystyle{\rm d}q}
                 {\displaystyle\sqrt{2E-V^2(q)}}\,-t\Bigg)
      \exp\Bigg({\rm i}\int\!\frac{\displaystyle U(q)\,{\rm d}q}
                 {\displaystyle\sqrt{2E-V^2(q)}}\Bigg)
      \nonumber\\
    &+&\bar\psi\phi_2\Bigg(\int\!\frac{\displaystyle{\rm d}q}
                 {\displaystyle\sqrt{2E-V^2(q)}}\,-t\Bigg)
      \exp\Bigg(-{\rm i}\int\!\frac{\displaystyle U(q)\,{\rm d}q}
                 {\displaystyle\sqrt{2E-V^2(q)}}\Bigg).
      \nonumber
\end{eqnarray}
 Using the Jakobi theorem (\ref{Jakobi})
\[
 \partial S /\partial \phi_1=const,\quad
 \partial S /\partial \phi_2=const,
\]
we obtain solutions for the odd Grassmann variables:
\begin{equation}
 \label{solution1}
   \psi=\psi_0\exp(-{\rm i}\int\!U(q(\tau))\,{\rm d}\tau),\quad
   \bar\psi=\bar\psi_0\exp ({\rm i}\int\!U(q(\tau))\,{\rm d}\tau).
\end{equation}
 Let us introduce the following series for the even Grassmann variable:
\begin{equation}
 \label{series}
   q(t)=x_{qc}(t)+q_0(t)\bar\psi\psi
       =x_{qc}(t)+q_0(t)\bar\psi_0\psi_0.
\end{equation}
 Then, from the Jakobi theorem we have:
\begin{equation}
 \label{solution2}
  -\frac{\partial S}{\partial A}
  =\int\!\frac{{\rm d}q}{\sqrt{2E-V^2(q)}}\,\bar\psi_0\psi_0-t\bar\psi_0\psi_0=const.
\end{equation}
 From (\ref{series}) and (\ref{solution2}) we can write
\begin{equation}
 \label{solution3}
  \int\!\frac{{\rm d}x_{qc}}{\sqrt{2E-V^2(x_{qc})}}-t=const.
\end{equation}
 Let us now evaluating the derivative
$\displaystyle\partial S /\displaystyle\partial E=const$.
 Taking into account the result (\ref{solution3}) and the following
expansions
\begin{eqnarray}
\label{expansion}
  U(q)=U(x_{qc})+U'(x_{qc})q_0\bar\psi_0\psi_0,\nonumber\\
  V^2(q)=V^2(x_{qc})+2V'(x_{qc})V(x_{qc})q_0\bar\psi_0\psi_0,\\
  f(V^2(q))=f(V^2(x_{qc}))+f'(V^2(x_{qc}))2V'(x_{qc})V(x_{qc})q_0\bar\psi_0\psi_0,
  \nonumber
\end{eqnarray}
we obtain:
\begin{equation}
 \label{solution4}
 \int\!\frac{{\rm d}q_0}{\sqrt{2E-V^2(x_{qc})}}
  =\int\!\frac{\big[A-U(x_{qc}(\tau))-V(x_{qc}(\tau))V'(x_{qc}(\tau))q_0(\tau)\big]}
   {2E-V^2(x_{qc}(\tau))}{\rm d}\tau .
\end{equation}
 This result can be presented in the form:
\begin{equation}
 \label{end}
  q_0(t)=\frac{\dot x_{qc}(t)}{\dot x_{qc}(0)}
  \left[q_0(0)-\int\limits_{0}^{t}\!{\rm d}\tau
  \frac{F-U(x_{qc}(\tau))}{2E-V^2(x_{qc}(\tau))}\right].
\end{equation}
 The obtained result coincides with the result obtained from the Lagrangian
equations of motion \cite{Junker1}.

 Thus the Hamilton-Jakobi equation and Jakobi theorem are
presented in Grassmann algebra.
 The action for the classical system characterized  by the SUSY Lagrangian
is presented in the explicit form.
 The results obtained using the Hamilton-Jakobi method coincide with
ones obtained previously from the Lagrangian equations of motion.

Acknowledgements.

I am very grateful to V.M.Tkachuk for comments and discusions.

$$ $$
 Tabunshchyk K.V. e-mail: tkir@icmp.lviv.ua

\end{document}